# A CROSS-LAYER APPROACH FOR MINIMIZING INTERFERENCE AND LATENCY OF MEDIUM ACCESS IN WIRELESS SENSOR NETWORKS

Behnam Dezfouli[1], Marjan Radi[2], Shukor Abd Razak[3]

Faculty of Computer Science and Information Systems,
Universiti Teknologi Malaysia (UTM), Malaysia
[1]DEZFOULI@IEEE.ORG, [2]RADI@IEEE.ORG, [3]SHUKORAR@UTM.MY

## ABSTRACT

*In low power wireless sensor networks, MAC protocols usually employ periodic sleep/wake schedule to reduce idle listening time. Even though this mechanism is simple and efficient, it results in high end-to-end latency and low throughput. On the other hand, the previously proposed CSMA/CA based MAC protocols have tried to reduce inter-node interference at the cost of increased latency and lower network capacity. In this paper we propose IAMAC, a CSMA/CA sleep/wake MAC protocol that minimizes inter-node interference, while also reduces per-hop delay through cross-layer interactions with the network layer. Furthermore, we show that IAMAC can be integrated into the SP architecture to perform its inter-layer interactions. Through simulation, we have extensively evaluated the performance of IAMAC in terms of different performance metrics. Simulation results confirm that IAMAC reduces energy consumption per node and leads to higher network lifetime compared to S-MAC and Adaptive S-MAC, while it also provides lower latency than S-MAC. Throughout our evaluations we have considered IAMAC in conjunction with two error recovery methods, i.e., ARQ and Seda. It is shown that using Seda as the error recovery mechanism of IAMAC results in higher throughput and lifetime compared to ARQ.*

## KEYWORDS

*Wireless Sensor Networks, MAC, IAMAC, Tree-Based Routing, Cross-Layer Optimization, Interference.*

## 1. INTRODUCTION

Wireless communication in wireless sensor networks are investigated in [1][2][3][4] and the irregularity and unreliability of low power wireless links are demonstrated. Accordingly, three distinct reception regions can be identified in a wireless link: connected, transitional, and disconnected. Since most of the links to neighboring nodes fall into the transitional region and they exhibit high variations in their quality (due to environmental noise, inter-node interference, etc.), there are many non-perfect links that may be selected by routing algorithms. Even though link estimation can help to select the best next hop neighbor along the path to the sink, nevertheless, concurrent transmissions of neighboring nodes cause immediate variations in link quality and lead to inter-node interference. Inter-node interference and high packet corruption rate result in more energy consumption per node, which is in contrast with long lifetime of tiny, low power sensor nodes. The current routing algorithms [3][5] and MAC protocol collision avoidance methods [6][7][8][9] cannot handle the effects of inter-node interference completely. At the MAC layer, to avoid collision in S-MAC [6], all the nodes that overhear control packets (i.e., RTS and CTS) are prevented from packet transmission. However, due to the multi-hop nature of packet transmission in wireless sensor networks, this mechanism results in very high end-to-end latency of this protocol. To improve this deficiency of S-MAC, Adaptive S-MAC [7] proposes adaptive node activation mechanism based on the estimated transmission duration between two neighboring nodes. In Section 2, we show that this mechanism outcomes in low lifetime and poor scalability of this protocol.

On the other hand, studies on protocol design for wireless sensor networks revealed the fact that reducing power consumption cannot be handled completely in one layer of the protocol stack and without any interaction with other layers [10][11][12]. However, as stressed in [11], cross-layer interaction and optimization must be performed with respect to the architecture. Neglecting architectural issues may result in unwanted interactions between components and hardens the understandability and improvement of protocols in the future.

In this paper, we propose an Interference Avoidance MAC protocol (IAMAC) that its main objective is to provide higher network lifetime by avoiding inter-node interference. In addition, IAMAC does not compromise delay as S-MAC does and provides lower end-to-end latency compared to S-MAC. Through information sharing between the MAC and network layer, proper decisions can be made at each node to avoid inter-node interference, while the delay is also reduced. Furthermore, since IAMAC is a cross-layer protocol, we show that IAMAC can be implemented with SP [13] architecture to perform its inter-layer interactions.

As the main applications of IAMAC we can consider monitoring and surveillance, i.e., nodes sample their environment periodically and send their results toward the sink node. Primary demands of these applications are long network lifetime, transmission reliability, and an acceptable latency, depending on the application. Usually, the delay of several minutes can be tolerated [14]. It should be noted that other applications can also be envisioned for this protocol.

The rest of this paper is organized as follows. In Section 2, by evaluating Adaptive S-MAC protocol, we give some motivations towards the design of IAMAC. Section 3 presents the proposed medium access control protocol. Performance evaluation is performed in Section 4. Some issues regarding the architecture and implementation of IAMAC are provided in Section 5. We conclude and provide directions for future works in Section 6.

## 2. Drawbacks of the Adaptive S-MAC Protocol

S-MAC [6] and Adaptive S-MAC [8] are two sleep/wake MAC protocols that are mainly designed for wireless sensor networks. Since energy efficiency and latency are two important evaluation metrics for MAC protocols, we investigate these two protocols in terms of these metrics. To reduce packet collision in S-MAC all the nodes which overhear control packets (i.e., RTS and CTS) are prevented from packet transmission. Although this mechanism results in low energy consumption, it also results in very high delay. To remedy this problem, Adaptive S-MAC is proposed and uses adaptive node activation based on the estimated transmission duration between neighboring nodes. Despite the lower latency of Adaptive S-MAC compared to S-MAC, this protocol has two main drawbacks: First, when two nodes communicate, neighboring nodes must be informed of the approximate duration of communication by overhearing RTS and/or CTS packets. Therefore, neighboring nodes can wake up before their predefined scheduled time to transmit their data packets. If so, data packets may incur lower delay because they may travel more than one hop in each frame (here, as proposed in [6], we used the term *frame* as a complete cycle of listen and sleep; later, we will introduce another term that better matches the concept of frame in sleep/wake MAC protocols). However, in low power wireless sensor networks, link qualities are so variable and this results in imprecise calculation of transmission duration, which also brings about longer idle listening and higher energy consumption per node. This additional energy consumption depends on the factors such as average number of neighboring nodes per node, radio type, and environment. Among these factors, neighbor count really limits the scalability of this protocol. For example, as the number of the nodes, which overhear control packets and adaptively wake up increases, the energy consumption of the network raises. Additionally, this increment in energy consumption also depends on radio switching and channel sampling costs. Moreover, when the environment affects the radio links to be more unstable, neighboring nodes will suffer longer idle listening durations due to their inaccurate wake up times. To measure the effects of network density on

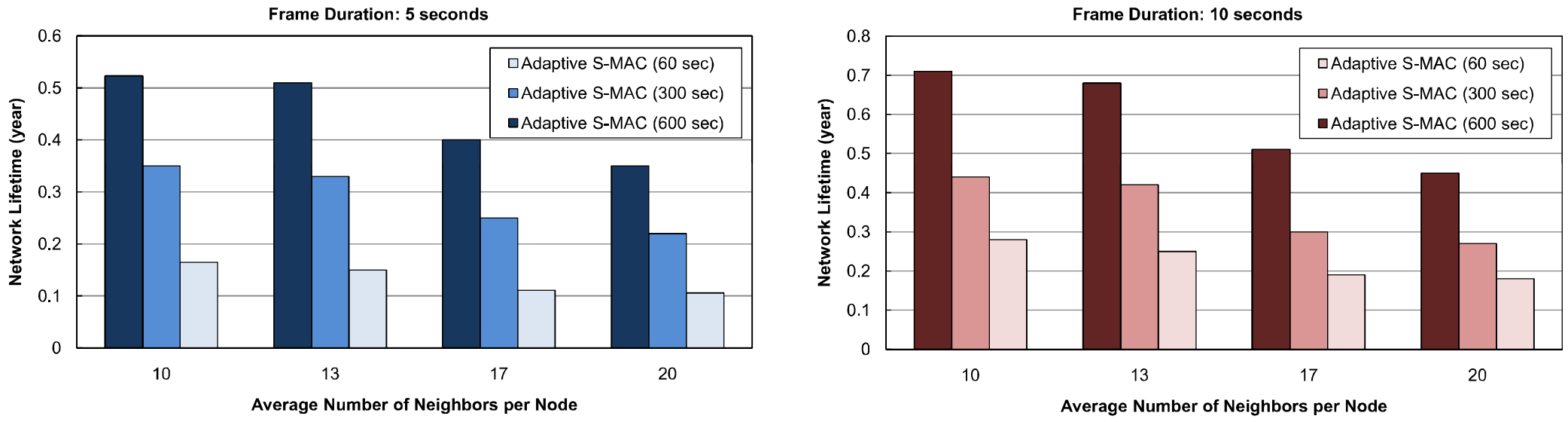

(a) (b)

Figure 1. Variations of the Adaptive S-MAC's lifetime versus the average number of neighbors per node. The value in each parenthesis indicates the packet generation interval at each node.

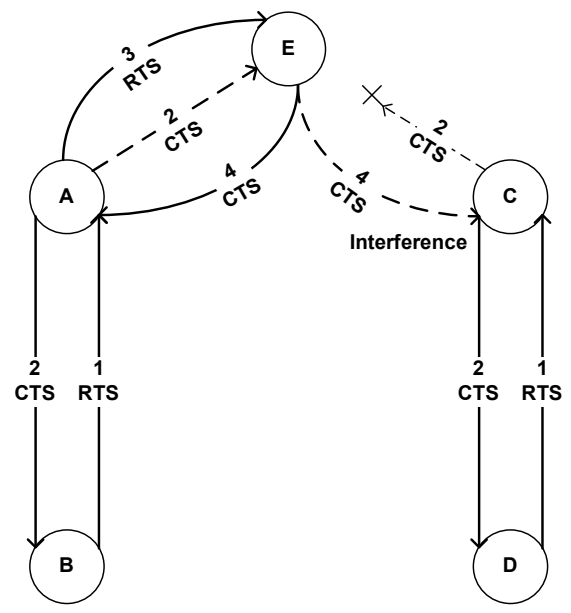

Figure 2. A sample scenario for interference in Adaptive S-MAC. The number on each arrow shows the time sequence. At time 1, node B and node D send RTS to node A and node C, respectively. When node A accepts data reception by sending CTS to node B, node E overhears this packet and captures the communication duration between node A and node B. Moreover, node E cannot overhear the CTS packet transmitted from node C to node D. Therefore, after the communication between node A and node B finishes, node E wakes up and interferes with data reception at node C.

Adaptive S-MAC we evaluated the lifetime of this protocol against the average number of neighbors per node in Figure 1. Our general simulation settings for the simulations of this paper are described in Section 4. According to Figure 1, Adaptive S-MAC is not scalable, i.e., as the average number of neighbors per node increases, the network lifetime decreases. This is the effect of increase in the number of the nodes, which adaptively wake up according to the communication duration between two neighboring nodes.

The second major problem of Adaptive S-MAC is the possibility of severe interference among neighboring nodes. A sample scenario is demonstrated in Figure 2 where node E interferes with data reception at node C.

According to these two disadvantages, Adaptive S-MAC provides very low network lifetime and it cannot be used in many applications. Motivated by these challenges, we try to propose a new MAC protocol to provide both the benefits of S-MAC and Adaptive S-MAC, i.e., long lifetime and low delay.

## 3. THE PROPOSED MAC PROTOCOL

In this section we introduce our proposed cross-layer MAC protocol. However, before proceeding to the MAC protocol description, we first introduce the routing algorithm.

### 3.1. Routing Protocol

We use spanning tree optimization as the routing algorithm. In the first step, each node broadcasts a fixed number of control packets and records the number of successfully received packets from its neighbors. After this step, each node's neighbor table includes link quality to neighboring nodes according to the ETX [5] link cost function. In the second step, sink node sets its cost to zero and broadcasts it to its neighbors. This broadcast operation is performed by transmitting Synch/Routing packets. A lightweight time synchronization protocol (such as [15][16]) is also used to synchronize sleep/wake schedule among the nodes. Thereby, a Synch/Routing packet includes time synchronization information along with the ETX cost to the sink. Upon receiving the Synch/Routing packet, each node adds the received cost to the link cost of the node from which this packet has been received. For example, consider node *A* receives a Synch/Routing packet from node *B*. Node *A* adds the cost contained in Synch/Routing packet to the cost of link *A*-*B*. If the resulting cost is less than the current cost of node *A* to the sink, node *B* will be selected as node *A*'s parent. Once the network reaches to a stable condition, each node follows its sleep/wake schedule (the sleep/wake structure of IAMAC will be explained in Section 3.2). Notice that the broadcast interval of Synch/Routing packets during the normal network operation depends on the time synchronization accuracy and route change frequency.

### 3.2. MAC Protocol

Similar to some of the previously proposed MAC protocols for wireless sensor networks [6][7][8][9][17], IAMAC uses sleep/wake scheduling for power conservation. However, its sleep/wake structure differs from the sleep/wake structure of S-MAC protocols so that it provides higher flexibility. Referring to the accuracy of time synchronization protocols for wireless sensor networks [15] [16], we assume that the maximum interval between consecutive time synchronizations cannot exceed 12 seconds. Consequently, in order to separate time synchronization from sleep/wake duration we proposed *Time Frame* and *Super Frame* sleep/wake structures. These structures are demonstrated in Figure 3. Since the Super Frame structure provides lower duty cycle, it can be used for lifetime critical applications. When the network lifetime and delay are equally important, Time Frame structure can be applied.

Regarding Figure 3, the first slot (i.e., Synch/Routing Slot) is dedicated to the transmission and reception of Synch/Routing packets (as described in Section 3.1). The next two slots (i.e., RTS Slot and CTS Slot) are devoted to the transmission and reception of RTS and CTS packets, respectively. Before introducing the proposed algorithms for RTS Slot and CTS Slot, we provide some details regarding the medium access mechanism in each of these slots. Figure 4 shows the channel access mechanisms during RTS Slot and CTS Slot. The RTS Slot is divided into mini slots called *RTS Contention Slot*. To transmit a RTS packet, a node must select a random RTS Contention Slot and random back off time for carrier sensing during the randomly selected mini slot. When the sender nodes (i.e., RTS transmitters) are within each other's carrier sensing range, random back off duration at the start of the selected RTS Contention Slot resolves contention among these nodes. As it is shown in Figure 4, when the node arrives at the forth RTS Contention Slot it performs carrier sensing for a random duration and then transmits its RTS packet. (Notice that all of the nodes, even if they don't want to transmit a RTS packet, should listen to the channel during the RTS Slot until the RTS Slot finishes or they become deactivated.) In the CTS Slot, when two or more nodes send their CTS packet simultaneously, none of them can be aware of the other's transmission and it may cause severe interference and packet loss during data transmission. Therefore, it is essential to avoid concurrent transmission of CTS packets. Furthermore, since the contention for CTS transmission is much lower than the contention for RTS transmission, CTS Slot is not divided into mini slots. Instead, each node should listen to the channel during a random back off time before transmitting its CTS packet.

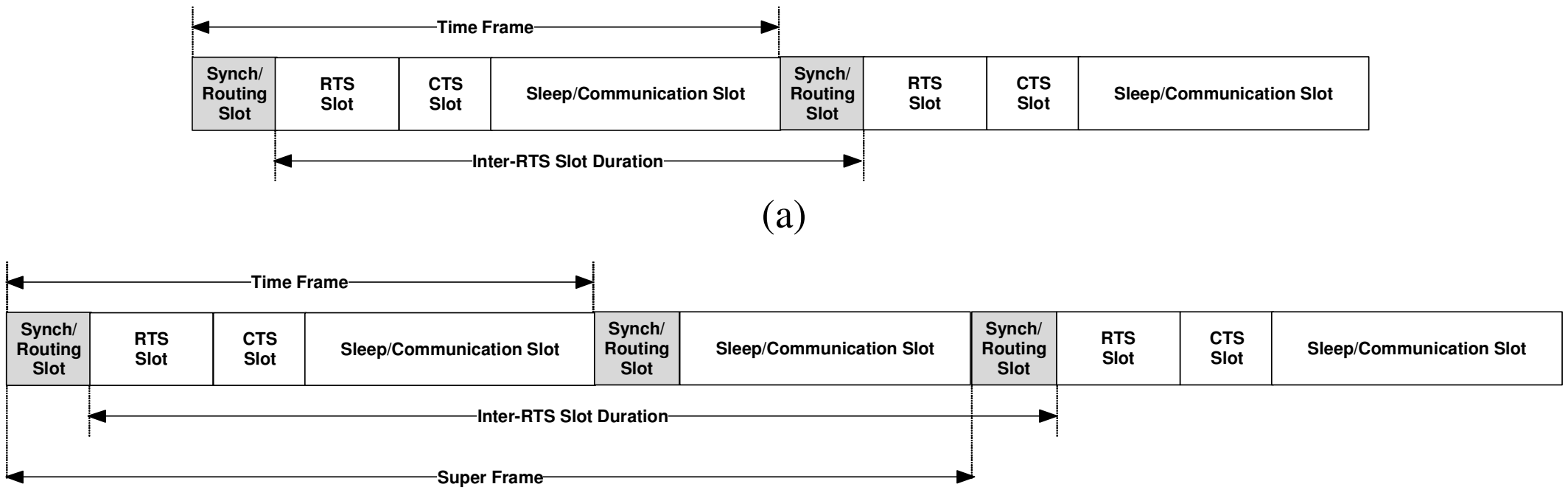

Figure 3. Time Frame and Super Frame structures. (a): If the duration between two consecutive RTS Slots is less than 12 seconds Time Frame structure can be used. (b): If the duration between two consecutive RTS Slots is more than 12 seconds Super Frame structure must be applied. When using Super Frame structure, Time Frame duration should not exceed 12 seconds.

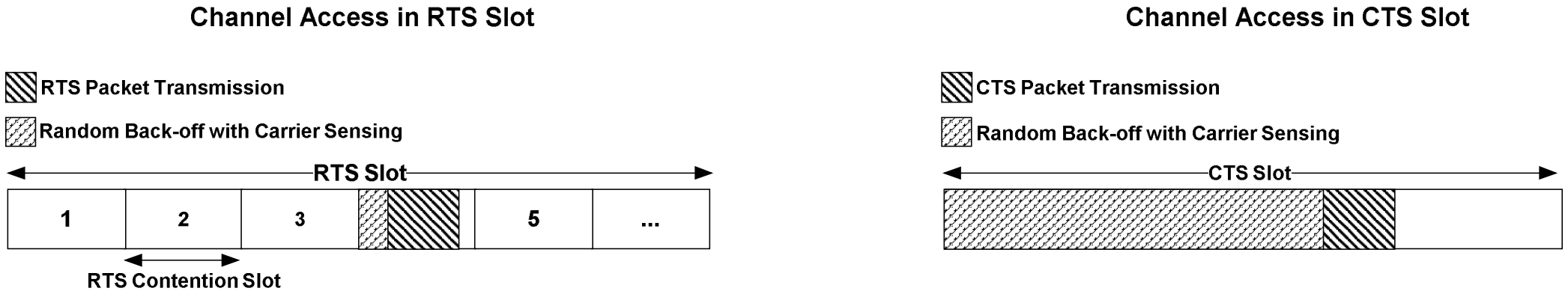

Figure 4. Channel Access Mechanisms During RTS Slot and CTS Slot

In this way, the probability of concurrent transmission of CTS packets will be very small due to the trivial signal propagation delay in wireless sensor networks.

IAMAC provides two different algorithms for RTS Slot and CTS Slot. Before explaining these algorithms, there are some points worth clarifying here. First, each node, in addition to the packet queue, must also include another queue to keep its received RTS packets during a Time/Super Frame; this queue is called *Received RTSs Queue*. Second, to prevent inter-node interference and control node operations, two Boolean control variables are defined: *CancelRTSTrans* and *CancelCTSTrans*. Third, in the proposed algorithms, node deactivation forces the node to go into sleep mode immediately.

Figure 5 demonstrates the flowchart of RTS Slot's algorithm. Because evaluating all the possible scenarios regarding to this algorithm is infeasible, we provide a somewhat simple scenario to clarify the operation of this algorithm. This scenario is depicted in Figure 6. The time below each step indicates the time progress as RTS packets are being received. Since each RTS packet can be received in a RTS Contention Slot, each time step corresponds to a RTS Contention Slot (notice that these time steps are not necessarily consecutive RTS Contention Slots). At Time 1, node *A* sends its RTS packet to node *B*. Upon reception of this packet, node *B* adds it to its Received RTSs Queue. When node *B* overhears a RTS transmission from node *E* to node *C* at Time 2, it conceives that replying to node *A* may result in inter-node interference. In addition, since the destination address of this overheard packet is the same as the parent address of node *B*, node *B* deletes the received RTS packet from its queue and changes its control variables to a new state. These new values for control variables allow node *B* to be a data transmitter (allowing it to send a RTS packet to its parent). Transmitting RTS packet from node *B* to node *C* at Time 3 does not change the state variables of node *E* because the overheard packet's destination address is the same as the parent address of node *E*. However, overhearing this RTS packet at node *D* changes its control variables to a new state since it can act only as a sender. At Time 4, node *D* sends its RTS packet to node *C*. Notice that overhearing this packet at node *B* does not change its control variables. At the end of the RTS Slot, node *C* includes RTS packets from *E*, *B*, and *D*; so it must respond to these packets during the CTS Slot. The respond mechanism can be performed in two ways: (1) using a single broadcast or (2) a CTS packet for each received RTS packet. Though using a single CTS packet is more energy efficient, incorrect packet reception at each node prevents that node from receiving its schedule for data transmission to its parent. In contrast, by transmitting multiple CTS packets we provide higher packet reception probability at the child nodes. In this paper, we use the maximum number of CTS transmissions, i.e., a CTS packet will be transmitted for each received RTS packet.

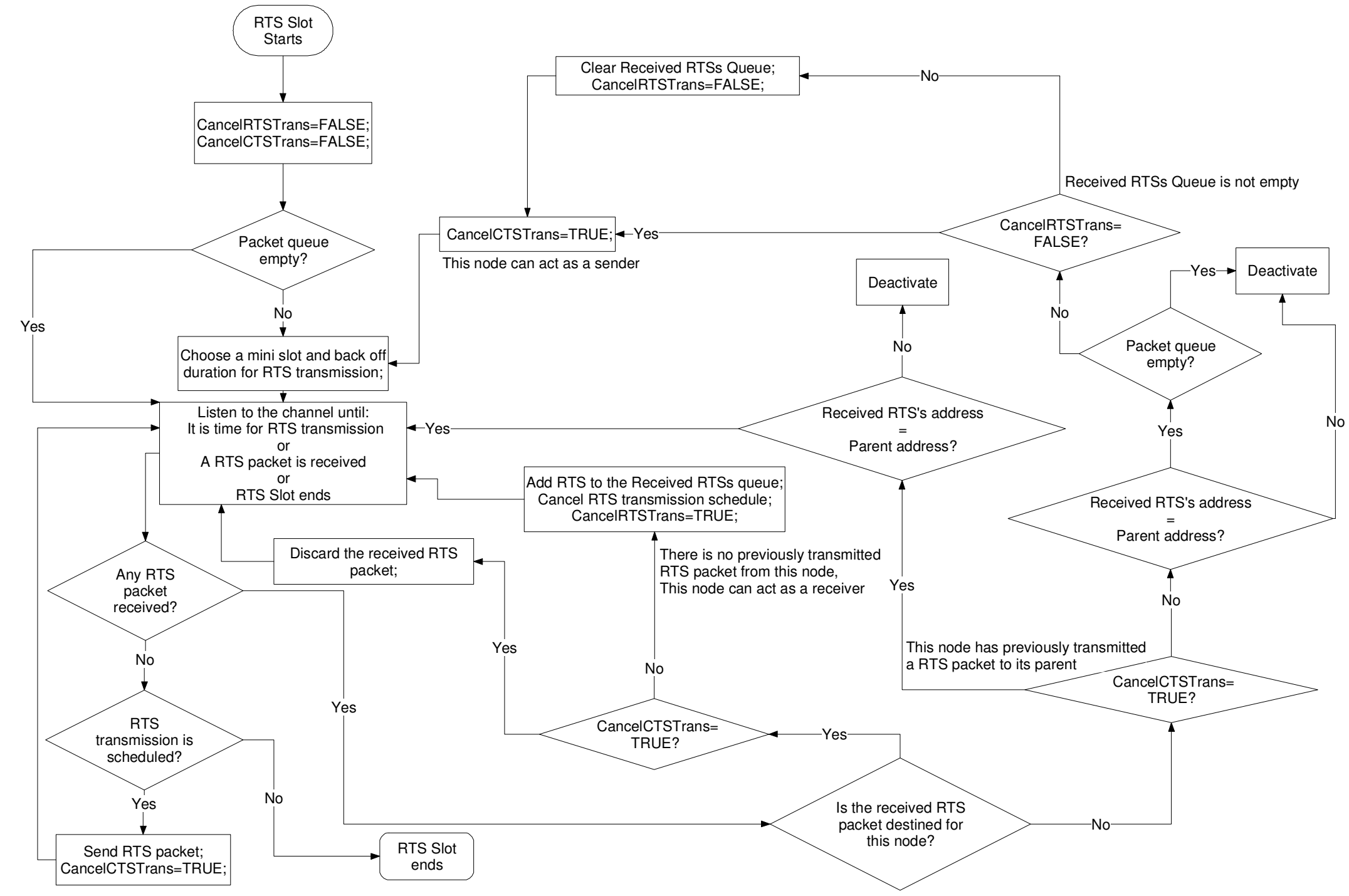

Figure 5. Flowchart of RTS Slot's Algorithm

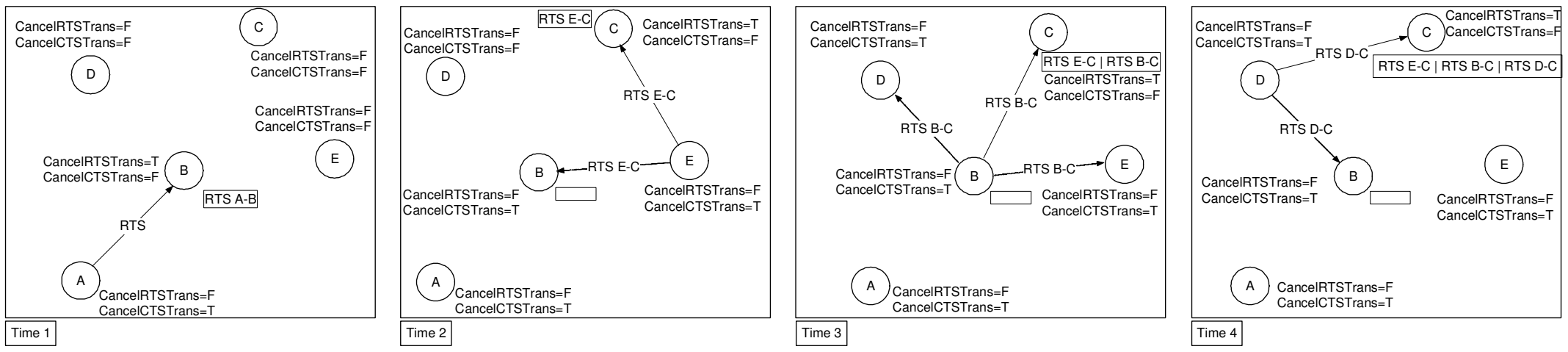

Figure 6. A Sample Scenario for RTS Slot

The second algorithm is the CTS Slot's algorithm, which acts as a complementary algorithm for RTS Slot's algorithm. Since the CTS Slot's algorithm is not complex and in order to provide some implementation details we provide it in the form of pseudo code. The corresponding algorithm can be seen in Algorithm 1.

Algorithm 1. Algorithm for CTS Slot

```
/*If the Received RTSs Queue is not empty and this node is allowed to send CTS
  packet:*/
If (ReceivedRTSsQueue.Length!=0)
    Choose a random time for CTS transmission;

While (CTS Slot is not finished)
{
 //When a CTS packet is received:
 If (a new packet is received)
    Pkt=Arrived Packet;

 If ( (CTS timer is reached) && (channel idle) )
    Send CTS packet;
    //This can be a single or multiple consecutive CTS transmissions

//Overhearing a CTS packet, cancel CTS transmission:
If ( (Pkt.RecAddress!=MyAddress) || (channel busy) )
    Deactivate;
    /*Due to link asymmetry and RTS packet corruption, we may receive a CTS
      packet that is not destined for this node. Therefore, in order to avoid
      interference, this node must be deactivated*/

/*If this node receives a CTS packet, it is allowed to transfer its data in
  Sleep/Communication Slot:*/
If (Pkt.RecAddress==MyAddress)
    Prepare for data transmission;
}
```

### 3.3. A Discussion on Slot Durations and Access Methods

According to the RTS Slot's algorithm, when a node reaches to its randomly selected RTS Contention Slot, a small back off time will be selected and the node continues listening to the channel. If nothing is sensed during this time, it can send its RTS packet. Otherwise, if the node receives a RTS packet destined for its parent, it selects another RTS Contention Slot among the remaining RTS Contention Slots and repeats these steps. If the received RTS packet is not destined for this node or this node's parent, the node becomes inactive. The required number of RTS transmissions in each RTS Slot depends on some factors such as Time/Super Frame capacity for data transmission, network traffic, and average number of children per node. If some nodes compete to grasp the channel and their RTS transmissions collide at the parent node, they will suffer more delay because they cannot transmit their data packets at the same Time/Super Frame. This situation occurs when the children cannot hear each other's transmission. Considering $n$ nodes with a common parent (i.e., $n$ children), when they cannot hear each other, the probability of correct reception of RTS packets from these nodes at the parent is as follows ($w$ is the number of RTS Contention Slots):

$$p_0 = \binom{w}{n} \times n! \times \left(\frac{1}{w}\right)^n \qquad (1)$$

Consequently, the RTS Slot duration depends on the average number of children per node. Furthermore, since the RTS Slot duration should be equal for all the nodes, scalability problems may appear. In order to remedy this problem, we can limit the maximum number of children per node. To this aim, when a node wants to select its parent, it also considers the number of children that its neighboring nodes currently have. Therefore, each node looks for a qualified node in terms of cost and number of children, and then selects that node as its parent.

Table 1. Default Simulation Settings

| Radio | | | |
|---|---|---|---|
| Modulation | FSK | Encoding | NRZ |
| Output Power | 0 dBm | Frame | 45 bytes |
| **Transmission Medium** | | | |
| Path Loss Exponent | 4 | $PL_{D0}$ | 55 dBm |
| Noise Floor | -105 dBm | $D_0$ | 1 m |
| **Other Parameters** | | | |
| Number of Nodes | 200 | Area | 100×100 m$^2$ |

Table 2. Detailed Parameters for ARQ and Seda

| Parameter | Symbol | Value |
|---|---|---|
| Maximum Packets per Frame | $MPF$ | Variable |
| Payload Length | $L_l$ | 29 |
| Physical and MAC Headers Length | $L_{phy_mac}$ | 16 |
| Packet Length | $L_P$ | 29+16 |
| Block Overhead | $L_{BO}$ | 2 |
| Block Length | $L_B$ | 29+2 |
| ACK Packet Length | $L_{ack}$ | 23 |
| Radio Speed (bps) | $S_R$ | 19200 |
| Bit Error Rate | $BER$ | Variable |
| Recovery Frame Overhead (byte) | $RF_{OV}$ | 5 |
| Sleep Duration (second) | $D_S$ | Variable |

## 4. EVALUATION

For precise evaluation of sensor network protocols, accurate modeling of wireless channel is of great importance. Accordingly, we implemented the link layer model from USC [1] in OMNeT++ framework. Then, IAMAC, S-MAC, Adaptive S-MAC, and spanning tree routing algorithm were implemented in separate modules. Table 1 represents our general simulation settings similar to the characteristics of MICA2 motes. The sink node is positioned at the middle of top edge. Table 2 provides more details regarding the data link layer parameters. Energy consumptions of radio and sensor operations are provided in [9]. In our evaluations, we may change some of these parameters with notification.

### 4.1. Interfering Nodes per Time/Super Frame

In this section, we evaluate the proposed protocol in the context of interference avoidance capability. In order to measure the interference avoidance level, we define $CS_{Ni}$ as the colliding set of node $N_i$ so that $CS_{Ni}$ is the number of nodes in the neighborhood of node $N_i$ which send their data packets concurrently with $N_i$ reception in the same Time/Super Frame. It should be noted that $CS_{Ni}$ excludes the node that is currently sending to $N_i$. It is evident that inter-node interference is possible when a node receives data packets while its $CS_{Ni}$ is not zero. Therefore, we sum the $CS_{Ni}$ value of the nodes (i.e., 200 nodes in our simulations) over the entire network in each Time/Super Frame. Figure 7 depicts this sum for two sizes of control packets (i.e., 18 and 28 bytes, except the physical and MAC layer headers). According to this figure, there are less than three interfering nodes in a 200-nodes network per Time Frame. For IAMAC, the only occasion in which $CS_{Ni}$ is not zero is due to the erroneous reception of control packets that is the effect of unreliable wireless communication. In addition, since smaller packets have lower error probability, less control packet size results in lower inter-node interference. As the sampling interval increases, the number of interfering nodes per Time/Super Frame reduces due to the lower contention for packet transmission. Although we cannot expect IAMAC to eliminate inter-node interference completely (because of unreliable wireless links and control packet corruption), these interferences have no severe effect on packet reception probability. Simulation results have demonstrated that for transitional region radius of 20 meters and control packet size of 18 bytes, about 70% of the interferer nodes reside 18 meters away from the receiver and about 97% of them reside 16 meters away from the receiver.

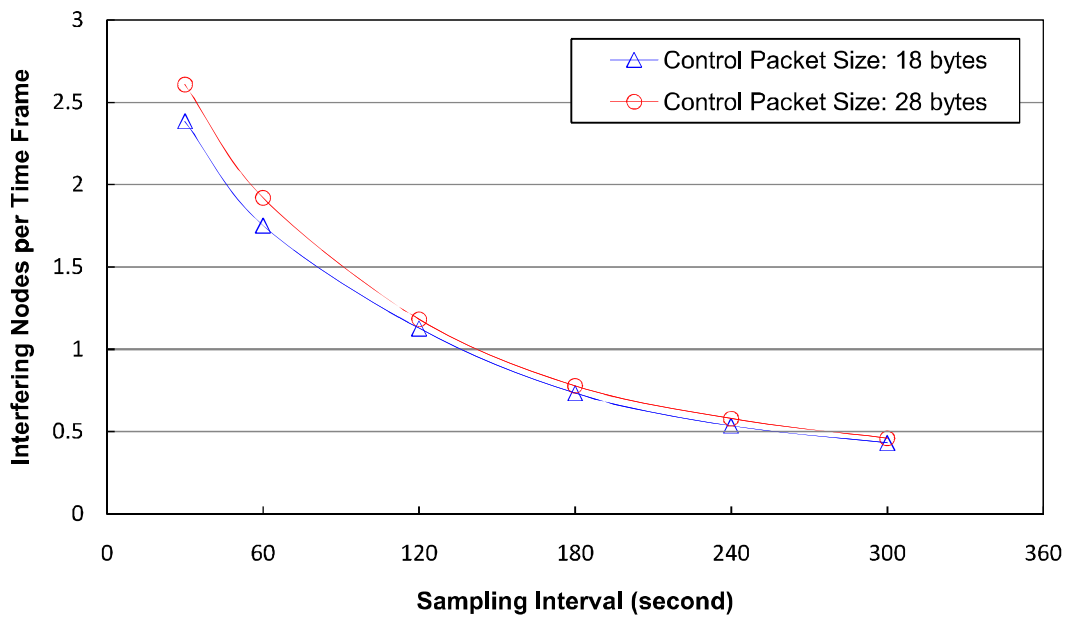

Figure 7. Interfering nodes per Time Frame versus sampling interval. (Time Frame Duration=1 sec.)

## 4.2. Throughput

One of the effective factors on network throughput is the number of concurrently transmitting nodes in each Time/Super Frame throughout the network. As the transitional region radius grows, IAMAC reduces number of the nodes, which concurrently transmit during a Time/Super Frame duration. The same effect can also be observed by increasing the output power level. Figure 8 demonstrates the average number of sender nodes during a Time Frame. Starting from 0 dBm, as the output power level reduces, due to less interference among the contending nodes the number of sender nodes per Time Frame increases. However, for each network density, this increment stops at a certain output power level. When the output power level goes below this threshold level, the average number of children per node decreases and therefore the number of sender nodes in each Time Frame reduces. The optimal output power level is inherently a cross-layer parameter that mainly depends on network density, routing protocol, and sampling rate. Notice that the output power levels less than -8 dBm caused the 100×100 $m^2$ network to be disjointed and therefore Figure 8 provides no simulation result for this situation.

For explicit evaluation of network throughput, we used two error recovery methods: ARQ and Seda [18], in conjunction with IAMAC. Seda is a novel error recovery technique that separates data framing from error recovery to achieve higher throughput. Due to two reasons we claim that Seda is more compatible with IAMAC, compared to ARQ. First, as we mentioned earlier, IAMAC reduces inter-node interference and packet corruption probability. If we use ARQ as IAMAC's error recovery mechanism, even for a properly received packet an ACK packet must be sent. This extra control packet transmission results in reduced network throughput and lifetime. In contrast with ARQ, Seda does not send any ACK packet and only considers lost packets in its error recovery mechanism. Second, the sleep/wake structure of IAMAC motivates us to use Seda. Since Seda uses long frames for data transmission, Sleep/Communication Slot provides enough time for this long frames to be transmitted. However, frame size is limited by the radio buffer capacity and in our simulations we considered 128 bytes for transmission buffer and 128 bytes for reception buffer [18].

Considering the sleep/wake cycling of IAMAC, analytical evaluation of ARQ and Seda can be performed as follows. To recover lost packets, ARQ retransmits unacknowledged packets after the timeout timer expires. Using Seda, after receiving a whole data frame (containing many blocks), in order to request for retransmission of corrupted blocks, a recovery frame will be sent to the sender node. Considering one retransmission per lost packet/block, maximum number of sent packets during a Time/Super Frame can be calculated as follows. Equation (2) shows this value for ARQ. Descriptions of parameters are provided in Table 2.

$$\begin{aligned}&\frac{L_P}{S_R}(MPF + MPF(1-(1-BER)^{L_P})) + \\ &\frac{L_{ack}}{S_R}(MPF + MPF(1-(1-BER)^{L_P})) + \gamma = D_S\end{aligned} \qquad (2)$$

Considering the payload portion of each block in Seda equal to the payload length in ARQ, maximum transmitted packets per Time/Super Frame can be computed using (3) for Seda.

$$\begin{aligned}&\frac{L_{phy_mac}}{S_R} + MPF.\frac{L_B}{S_R} + (1-(1-BER)^{(L_B.MPF)}). \\ &[\frac{RF_{OV}}{S_R} + MPF.\frac{(1-(1-BER)^{L_B})}{S_R} + \frac{L_{phy_mac}}{S_R} + MPF.(\frac{(1-(1-BER)^{L_B})}{S_R}.\frac{L_B}{S_R})] + \gamma = D_S\end{aligned} \qquad (3)$$

Notice that $\gamma$ is independent from bit error rate and depends on some of the characteristics of radio transmitter and node hardware such as the radio buffer size and processing speed. Nevertheless, due to the small value of $\gamma$ we neglected its effect in our analysis. Figure 9 shows these two equations against BER. As it can be observed, Seda provides more packet transmissions during a Time Frame. Decoupling framing from error recovery and short data blocks of Seda, makes Seda less vulnerable against wireless channel errors. In addition, Seda applies less physical and MAC layer overheads per payload. As the result, Seda experiences less packet corruption rate and can transmit more volumes of data during a Time/Super Frame. Notice that in Figure 9, since we have considered just one retransmission per corrupted packet/block, both curves tend to their fixed value by increasing BER.

In order to measure the maximum network throughput, we forced each node to sample the environment as fast as it can and transmit its data packets with maximum capacity. Figure 10 shows the throughput of IAMAC in combination with Seda and ARQ. According to this figure, IAMAC with Seda achieves higher throughput than ARQ, which also confirms our analytical results. In this figure, notice the rise and fall of the network throughput that is similar to the behavior observed in Figure 8.

## 4.3. Lifetime

Figure 11 shows the lifetime of IAMAC against different sampling intervals. Starting from 60 seconds, as the sampling interval increases the number of generated packets per node reduces and leads to higher network lifetime. Furthermore, as we increase the sampling interval, lifetime reaches to its maximum value at a specific point. At this point, a trade off is established between number of transmissions per Time/Super Frame, nodes' active duration, and number of deactivated nodes. Considering this situation, in addition to the large number of transmissions per Time/Super Frame, many nodes are also deactivated by overhearing control packets. If we increase the sampling interval and go beyond this point, number of concurrent transmissions per Time/Super Frame and number of deactivated nodes will be reduced and therefore lifetime slightly decreases. For short Time/Super Frame durations the average duty cycle of the nodes is high and the network can benefit from the effects of increased transmissions per Time/Super Frame and more node deactivations. In contrast, for long Super Frames the average duty cycle is inherently low and increasing sequential transmissions per Time/Super Frame or higher node deactivations cannot result in noticeable increase of lifetime. This behavior is also visible in Figure 12, in which the average duty cycle of the nodes is demonstrated.

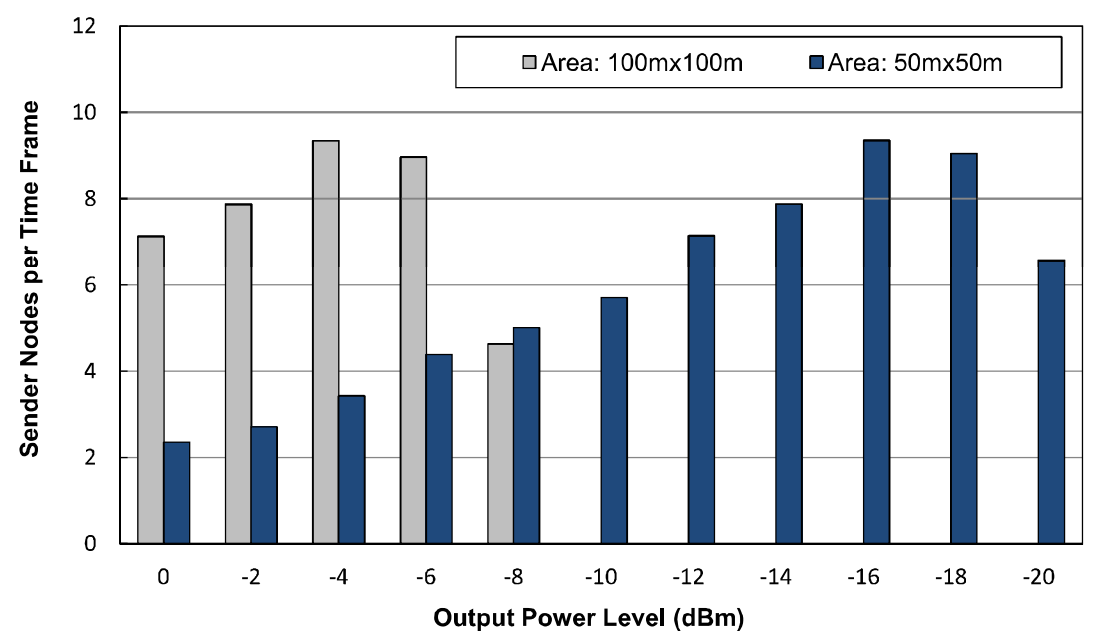

Figure 8. Average number of sender nodes per Time Frame. Each network density corresponds to an optimal output power level, which trades off between radio interference level and number of children per node. (Time Frame duration=1 sec, sampling interval=60 sec.)

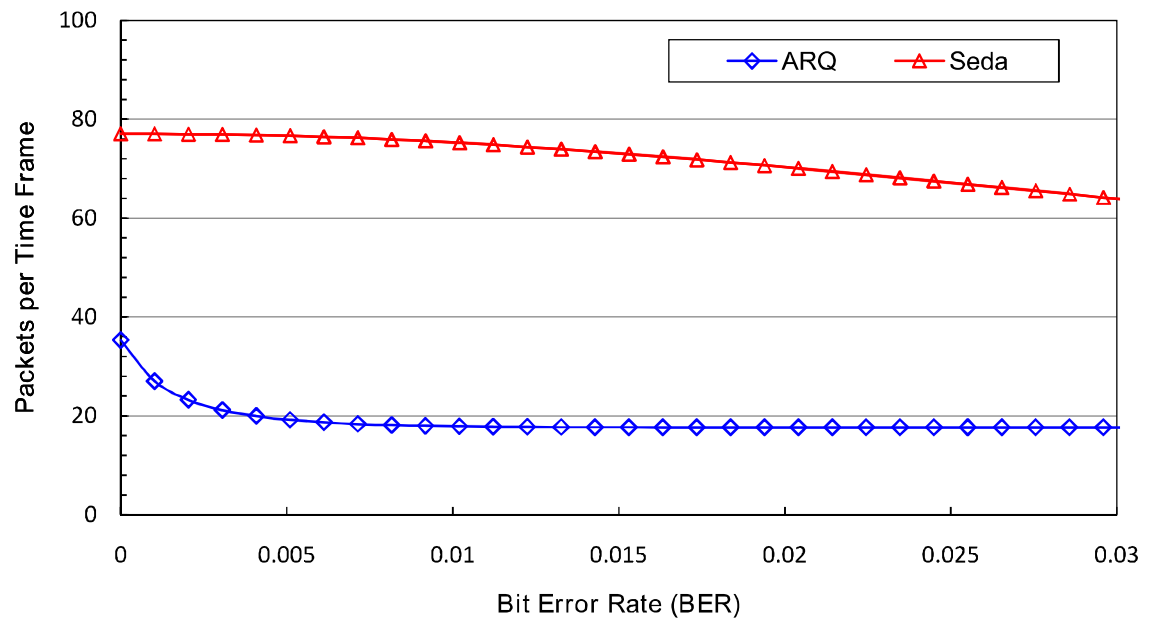

Figure 9. Number of data packets per Time Frame, considering one retransmission for every corrupted packet/block. (Time Frame duration=1 sec.)

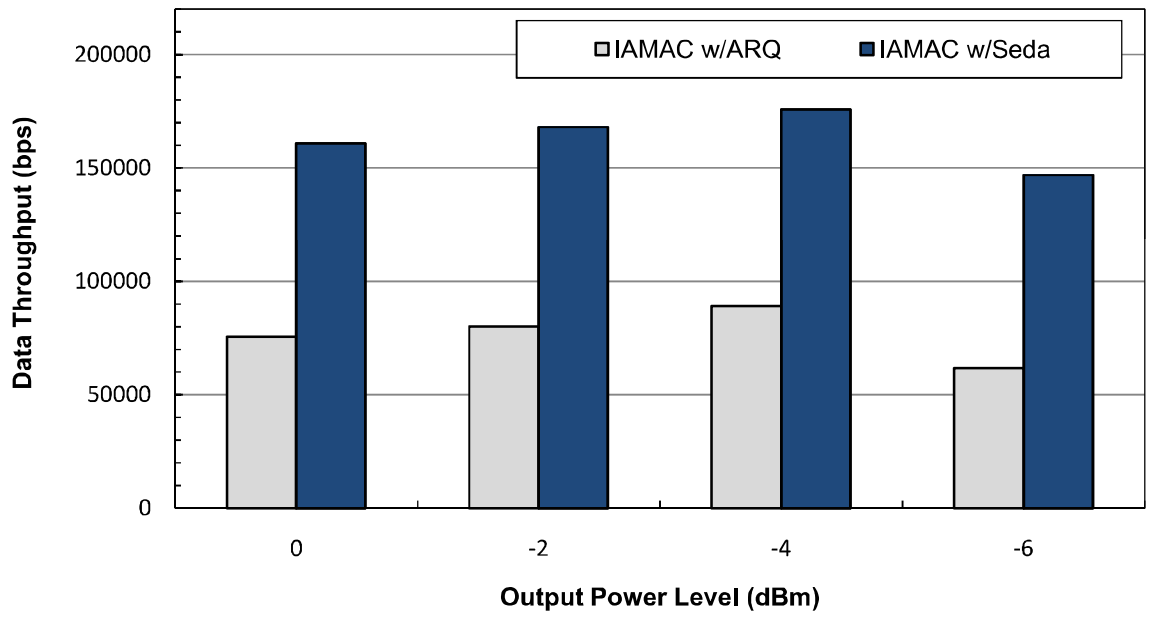

Figure 10. Effect of output power level on network throughput. (Time Frame duration=1 sec, sampling interval=1.1 sec.)

According to Figure 11, generally, as we increase the sampling interval, lifetime also increases. However, it should be noticed that by increasing the sampling interval, we cannot increase the lifetime indefinitely since: (1) each node has a limited initial energy (we have considered a 2400 mAh battery per node), (2) synchronization overhead limits the maximum network lifetime, (3) by increasing the Time/Super Frame duration number of queued packets per node increases and results in shorter sleep duration.

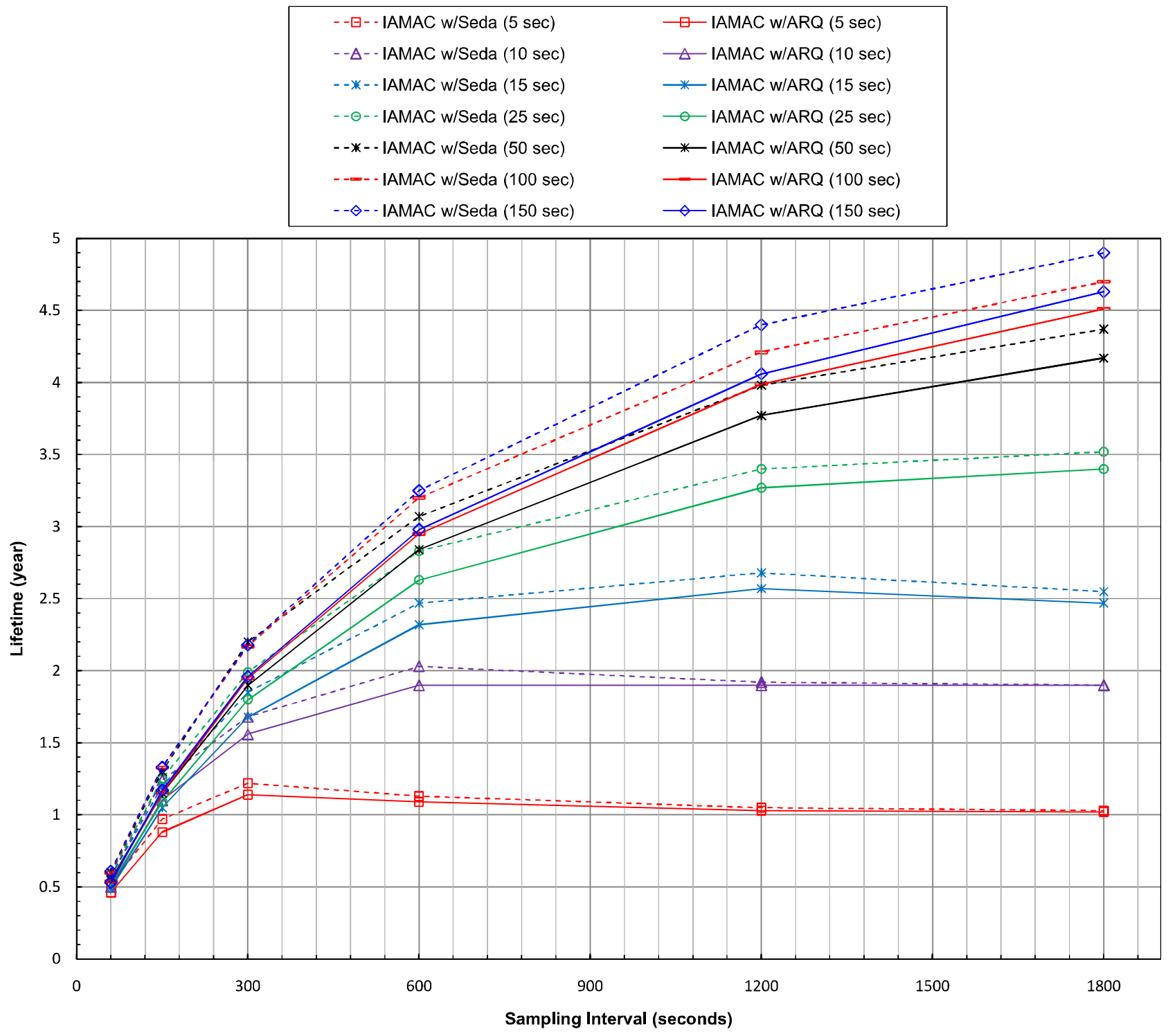

Figure 11. IAMAC's network lifetime as a function of sampling interval. The value in each parenthesis indicates the Time/Super Frame duration. As the sampling interval increases, the lifetime also increases because less time is spent on transmission and reception of data packets. Increasing Time/Super Frame duration also increases lifetime. This is due to the less overhead of active slots (i.e., Synch/Routing Slot, RTS Slot, and CTS Slot), compared to the whole Time/Super Frame duration. Also, Seda can improve the lifetime of IAMAC and this improvement is more evident for long sampling intervals and lengthy Time/Super Frame durations. When number of transmitted data packets in each Time/Super Frame is high, Seda can benefit from its low packet corruption rate and efficient error recovery.

Figure 13 demonstrates the lifetime of IAMAC against S-MAC and Adaptive S-MAC. It is evident that IAMAC provides significant increase in lifetime compared to Adaptive S-MAC. As discussed in Section 2, the lower lifetime of Adaptive S-MAC is mainly due to its adaptive listening mechanism. Even though with equal Time Frame durations IAMAC provides lower lifetime against S-MAC, it will be shown in Section 4.4 that IAMAC obtains higher performance than S-MAC in terms of lifetime and delay.

## 4.4. End-to-End Latency

In Figure 14, the latency of IAMAC is evaluated and compared to S-MAC and Adaptive S-MAC. Notice that the vertical axis is demonstrated in logarithmic scale to clear the differences between different curves. When IAMAC senses probable inter-node interference and prevents some nodes from data transmission, their data packets will experience a delay equal to the Time/Super Frame duration. At low sampling intervals, contention for channel access is very high and many nodes must be prohibited from communication. Nevertheless, in comparison with S-MAC, IAMAC provides lower delay. This is due to multiple transmissions to a common parent during a Time/Super Frame duration and its lower packet corruption rate.

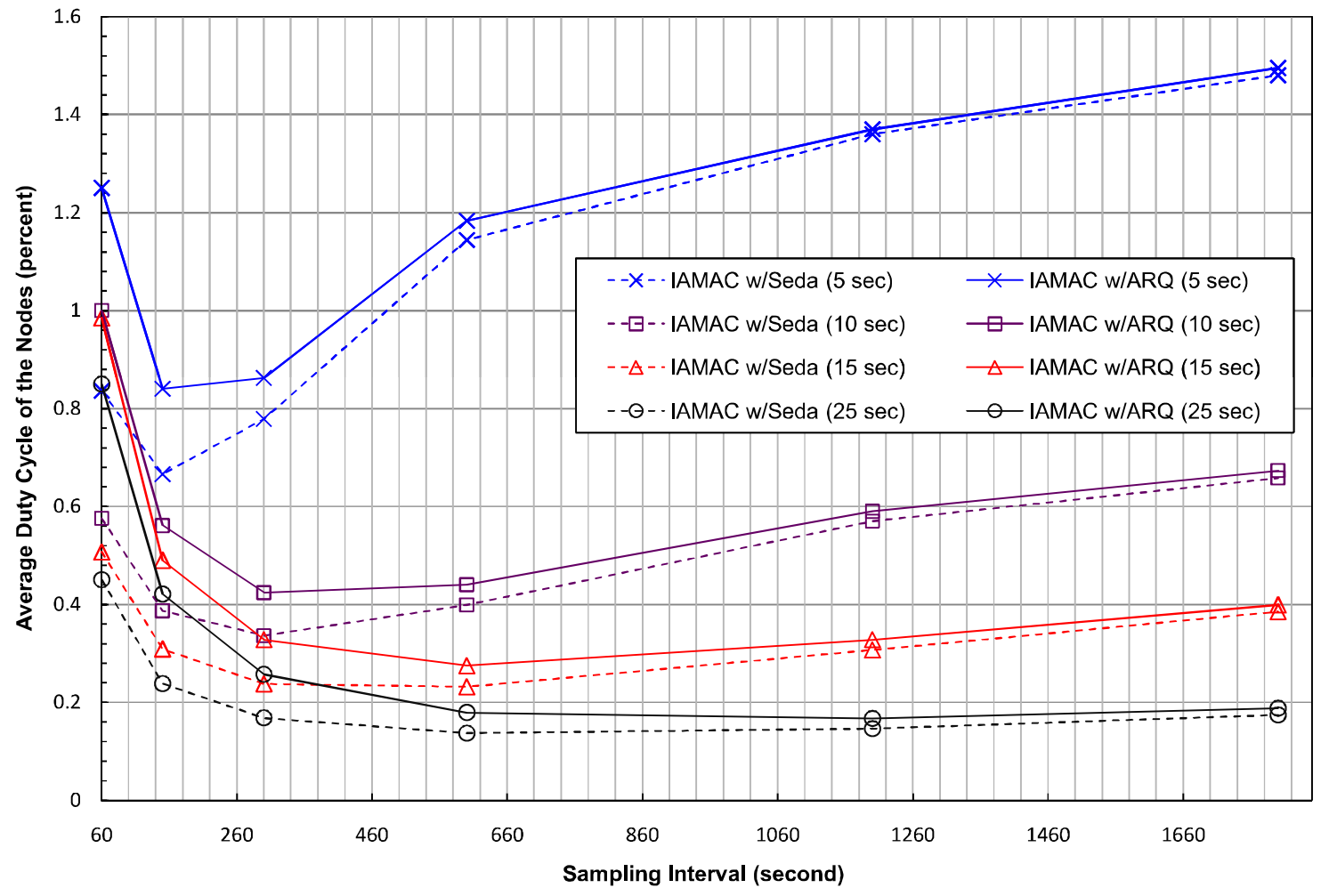

Figure 12. Variations of duty cycle against sampling interval. Notice the fall and rise of each duty cycle around a specific sampling interval. These minimum values for average duty cycle appear as the result of trade off between node active time, number of sequential transmissions per Time/Super Frame, and number of deactivated nodes. For long Time/Super Frame durations, the average duty cycle will be inherently low and this behavior is less evident.

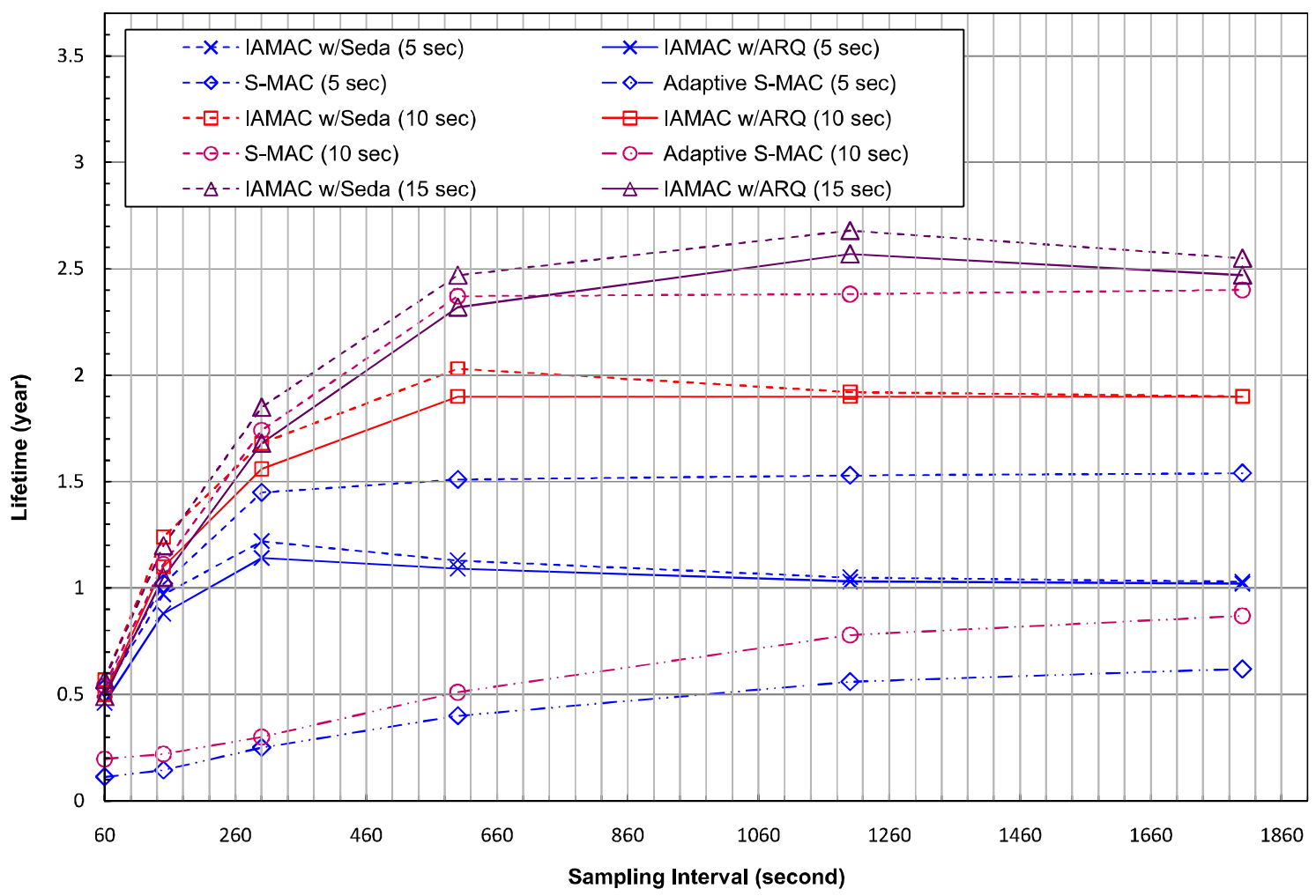

Figure 13. Network lifetime of IAMAC, S-MAC, and Adaptive S-MAC versus sampling interval. The value in each parenthesis demonstrates the Time/Super Frame duration for IAMAC and frame duration for S-MAC and Adaptive S-MAC.

As we have mentioned earlier, IAMAC provides higher performance than S-MAC in terms of lifetime and delay. For example, consider IAMAC (10 sec) and S-MAC (5 sec) in Figure 13. It can be seen that IAMAC provides higher lifetime than S-MAC. Furthermore, according to Figure 14, IAMAC (10 sec) has lower delay than S-MAC (5 sec). Consequently, IAMAC provides higher lifetime and lower delay compared with S-MAC.

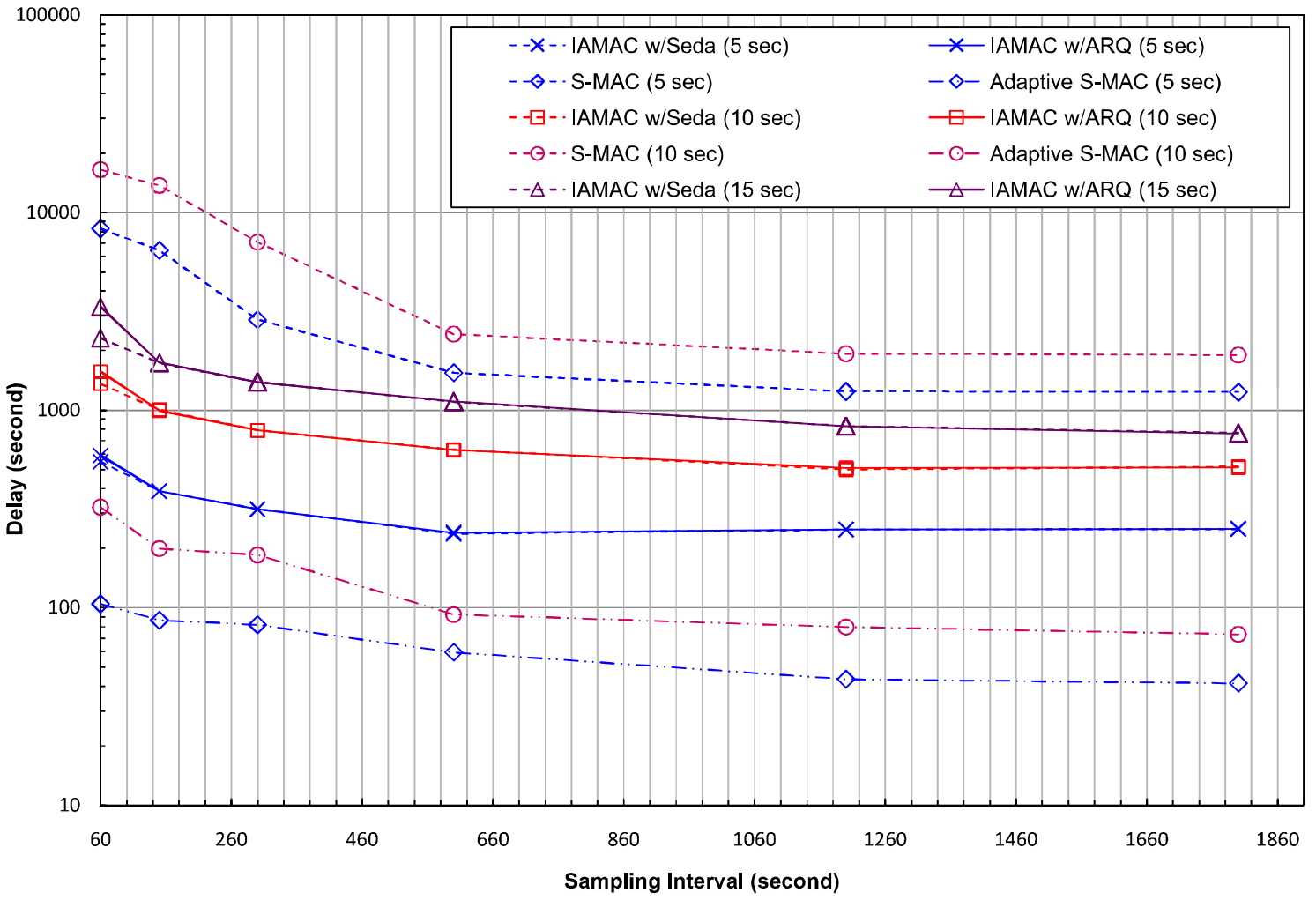

Figure 14: End-to-end delay of IAMAC, S-MAC, and Adaptive S-MAC. The value in each parenthesis demonstrates the Time/Super Frame duration for IAMAC and frame duration for S-MAC and Adaptive S-MAC.

### 4.5. Buffer Requirement Analysis

Since low power wireless sensor networks use multi-hop packet transmission, MAC protocols play an important role in per-hop data delivery. Furthermore, as far as the internal memory of sensor nodes is limited, the analysis of average packet queue size seems to be necessary. With our 200-nodes network, the average length of data packet queue versus different sampling intervals is demonstrated in Figure 15. According to this figure, memory demands can be fulfilled easily. For example, with 100 seconds of Super Frame duration, ARQ and Seda need to store about 690 and 340 packets per node, respectively. Therefore, if we consider 29-bytes payloads, 19.5 KB and 9.6 KB are needed for ARQ and Seda, respectively. This amount of memory can be provided by the internal memory of microcontrollers.

In Figure 15 observe that at each sampling interval, increment in the Time/Super Frame duration increases the average number of queued packets per node. This is due to higher number of generated packets per Time/Super Frame duration and increased contentions for medium access. Additionally, because of the lower overhead of Seda in packet transmission and recovery, it results in lower mean queue length than ARQ.

## 5. Architectural Issues

Although increasing inter-layer interactions in cross-layer optimization provides more opportunities for performance optimization, however, the effects of these interactions must be considered carefully. Establishing connections and interactions between different protocols may destroy system modularity and impede the understandability and optimization of the protocols [11][12][19]. To this aim, SP architecture [13] tries to provide richer inter-layer interactions while it also preserves modularity. In this architecture, through the SP abstract layer the upper and lower layers can communicate with each other. On the other hand, as we have seen before, IAMAC is based on the interactions of MAC and network layer. Accordingly, IAMAC can be implemented in the SP architecture in which the MAC and network protocol use the SP layer to perform their interactions. Figure 16 demonstrates the SP architecture containing IAMAC in its MAC layer. By integrating IAMAC and SP we can apply cross-layer optimization while we also maintain the modularity of the architecture. As a result, improvement of IAMAC or inclusion of other network protocols can be achieved easily in the future.

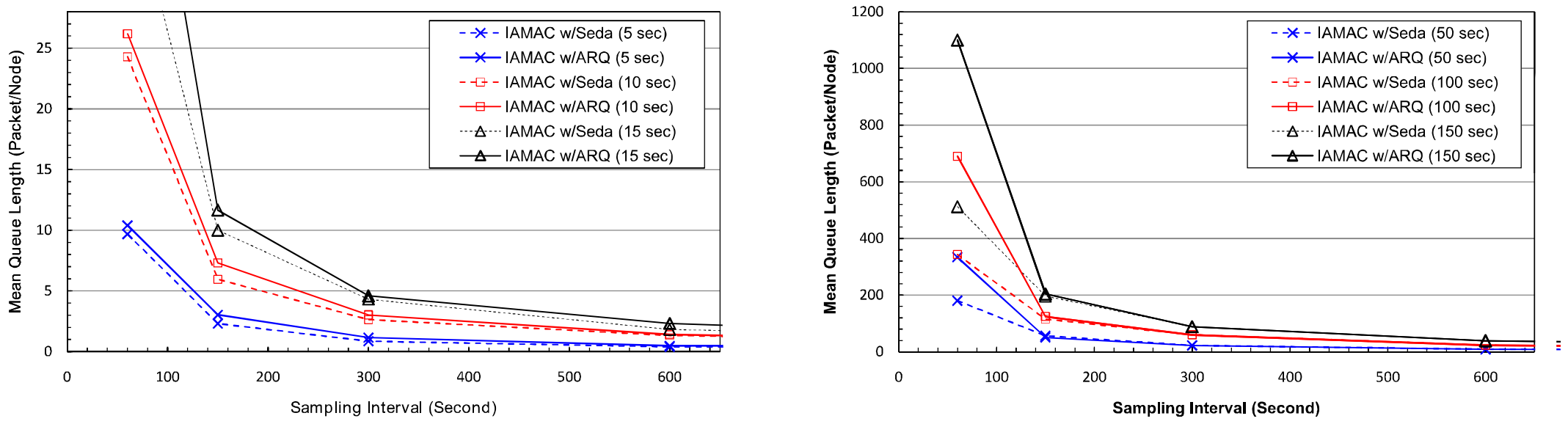

(a) (b)

Figure 15. Mean queue length per node with IAMAC as the MAC protocol. (a): Mean queue length for short Time/Super Frame durations (15 seconds and less). (b): Mean queue length for long Super Frame durations (50 seconds and more).

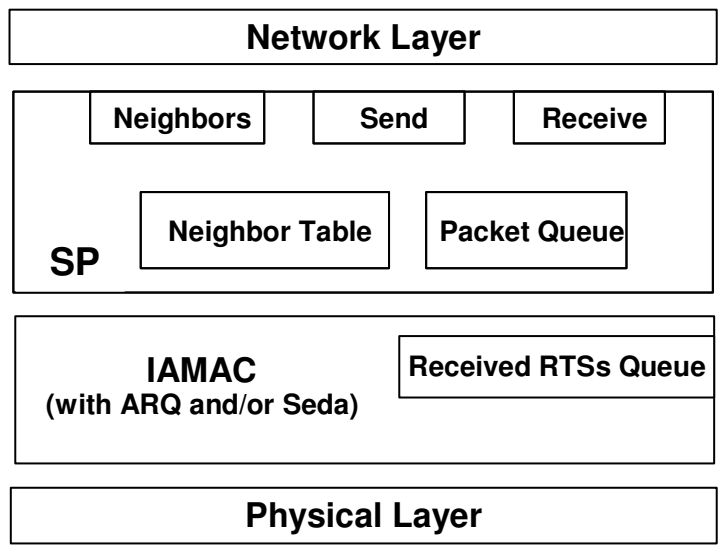

Figure 16. Implementing IAMAC within SP architecture. Network layer protocol and IAMAC can access to the Neighbor Table and Packet Queue data structures. Through three main operations of SP (i.e., Neighbors, Send, and Receive), neighbor table can be managed and data packets can be sent or received via the MAC protocol.

## 6. Conclusion

In this paper we proposed a novel medium access control protocol (IAMAC) to increase the performance of wireless sensor networks in terms of lifetime and delay. IAMAC achieves its high performance by three main mechanisms. First, IAMAC reduces inter-node interference and packet corruption via its two interference avoidance algorithms. Second, it utilizes the tree routing structure as multiple nodes can transmit to a common parent during a Time/Super Frame. This technique leads to lower control packet overhead and reduced per-hop latency. Third, IAMAC is a sleep/wake MAC protocol that separates Time/Super Frame duration from synchronization; therefore, it is possible to trade between lifetime and delay depending on application requirements.

Considering a realistic data link model, we conducted extensive simulations to evaluate the performance of IAMAC. According to the results, IAMAC provides higher lifetime compared to S-MAC and Adaptive S-MAC, while its end-to-end latency is less than S-MAC. Therefore, IAMAC can be an appropriate choice for lifetime critical applications such as surveillance and monitoring. In addition, due to its Time and Super Frame structures, IAMAC has more flexibility than S-MAC and Adaptive S-MAC. Finally, we showed that by implementing IAMAC into the SP architecture it can perform its inter-layer interactions through the SP abstract layer.

As demonstrated in the simulations, there are some parameters that affect the performance of IAMAC. For example, duration of contention slots (i.e., RTS Slot and CTS Slot) and transmission power highly affect network lifetime and latency. Even though simulation can be used to find optimal values for these parameters, it is difficult and time consuming. Accordingly, developing an analytical method for determining these optimal values can be useful. Node density, sampling rate, and some physical layer characteristics are among the input parameters of analytical model.

## Authors

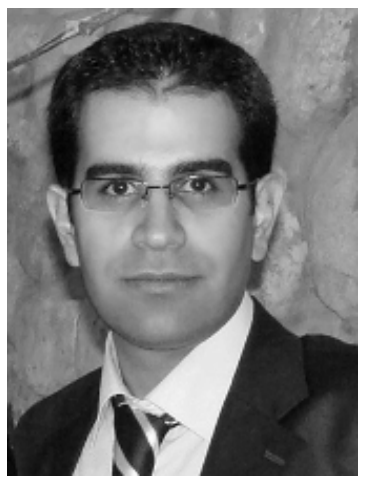

**Behnam Dezfouli** received his B.S. degree in hardware engineering and his M.S. degree in software engineering from Azad University, Iran, at 2006 at 2009, respectively. He served as a lecturer at Azad University and PN University, Iran, during 2007 to 2009. He is currently towards his Ph.D. degree in computer science at the Department of Computer Science and Information Systems, Universiti Teknologi Malaysia (UTM). His research interests include wireless ad hoc and sensor networks, cross-layer optimization techniques, mathematical modeling and performance analysis, and developing network simulation frameworks. He is also a student member of IEEE.

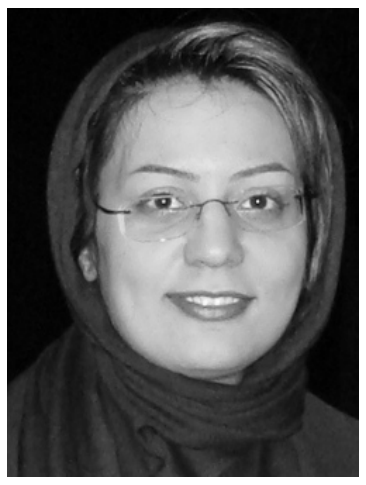

**Marjan Radi** received her B.S. and M.S. degrees both in software engineering from Azad University, Iran, in 2006 and 2009, respectively. Between 2007 and 2009 she was a lecturer with the Department of Computer Engineering, PN University and Azad University, Iran. She is currently working towards her Ph.D. degree in computer science at the Department of Computer Science and Information Systems, Universiti Teknologi Malaysia (UTM). Her research interests include routing algorithms, congestion control mechanisms, quality of service support, and resource allocation in wireless sensor and ad hoc networks. She is a student member of IEEE.

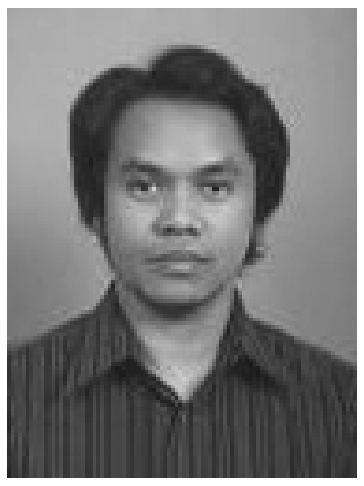

**Shukor Abd Razak** is a senior lecturer at the Universiti Teknologi Malaysia (UTM). His research interests are on the security issues of mobile ad hoc networks, mobile IPv6 networks, vehicular ad hoc networks, and network security. He is the author and co-author of many journal and conference papers at national and international levels.